\documentstyle{elsart}
\newcommand{\beq}{\begin{equation}}
\newcommand{\eeq}{\end{equation}}
\newcommand{\rOT}{(-1,2)\oplus (1,2)}
\newcommand{\rTT}{(-2,3)\oplus (2,3)}
\newcommand{\mmt}{(m,m+3)\oplus (m+3,m)}
\newcommand{\mms}{(m,m+6)\oplus (m+6,m)}
\newcommand{\cqa}{c_{1,q}^m}
\newcommand{\cqb}{c_{2,q}^m}
\newcommand{\cD}{{\cal D}}

\begin{document}
\begin{frontmatter}
\title{ On gauge--independence in quantum gravity}

\author{Dmitri V.Vassilevich}
\thanks{e.mail: vasilevich@phim.niif.spb.su}
\address{
Department of Theoretical Physics, St.Petersburg University, 198904
St.Petersburg, Russia}

\begin{abstract}
We prove gauge-independence of one-loop path integral for
on-shell quantum
gravity obtained in a framework of modified geometric approach. We
use projector on pure gauge directions constructed via quadratic
form of the action. This enables us to formulate the proof
entirely in terms of determinants of non-degenerate elliptic
operators without reference to any renormalization procedure.
The role of the conformal factor rotation in achieving
gauge-independence is discussed.
Direct computations on $CP^2$ in a general three-parameter
background gauge are presented. We comment on gauge dependence
of previous results by Ichinose.
\end{abstract}
\end{frontmatter}

\section{Introduction}

A selfconsistent approach to quantization of gauge theories was
suggested in pioneering paper by Faddeev and Popov \cite{A1} a quarter
of century ago. However, equivalence between different forms of
path integral is still under investigation (see, for example,
\cite{A2,A3,A4,A5,A6,A7,A8,A9}). In particular, it was
demonstrated \cite{A10} that under
certain conditions different gauge fixings can give inequivalent
results. There is also a problem specific for quantum gravity
\cite{A11} which is related to the choice of time and reduced phase
space quantization. Hence the problem of equivalence of
different forms of path integral in quantum gravity deserves
further investigation.

    Gauge-independence of the on-shell effective action in a
general gauge theory was demonstrated from different points of
view \cite{A1,FS,BV,Tu,Ka} (for a more complete bibliography
see recent monograph \cite{BOS}). This general statement was
supported by explicit calculations \cite{KTT} in one-loop
quantum gravity. Thus the problem was solved at least at
perturbative level.
However, recently some authors \cite{Ich,Pro}
claimed that due to non-renormalizability of quantum gravity
all proofs of gauge-independence are invalid. As far as
only the pure Einstein gravity was concerned, their
argumentation was based on a single computation by
Ichinose \cite{Ich}, which was not supported by
independent calculations \cite{LaR}. According to this point
of view, which was expressed in its extreme form in Ref.
\cite{Pro}, only physical quantities should be
gauge-independent. But no such quantities could be derived
from a non-renormalizable theory. Hence, the one-loop
on-shell effective action in quantum gravity is non-physical
and can be gauge-dependent. This dependence does not lead to
any real problem since, again, quantum gravity is
non-renormalizable and non-physical. From our point of view
this argumentation is at least incomplete. First, the term
"physical" in this context does not mean that a quantity
should be directly measurable in some experiment, it rather
means that a physical quantity is one which can be extracted
from on-shell effective action. Hence, the hand-waving
argumentation \cite{Pro} is not enough, one should find
a loop-hole in the proofs \cite{A1,FS,BV,Tu,Ka}
of gauge-independence.
The proof \cite{Tu}, though being quite formal, does not
rely upon renormalizability of a gauge theory.
Second, at one-loop order all gauge theories look
very much alike. The fact of non-renormalizability does
not show up at this order of perturbation theory.

    The aim of the present paper is to re-examine the situation
with gauge-independence on on-shell one-loop quantum gravity.

We suggest a new proof of gauge-independence for
a family of one-loop path integrals in quantum gravity. As in
the papers \cite{A2}, we use geometric
approach \cite{A12} to quantization
of gauge theories. However, we do not use integration over the
so-called adapted coordinates. Note, that these adapted
coordinates satisfy standard gauge conditions. We integrate
over variables satisfying arbitrary gauge conditions. Our path
integral can be viewed upon as Mazur and Mottola \cite{A13} path
integral in arbitrary gauge. This is the first point where our
technique is new. The second one is related to the
representation of projector onto the space of pure gauge fields
in terms of quadratic form of the action. A similar trick was
used before in curved space QED \cite{A14}. We shall demonstrate
equivalence of path integral in different gauges to the one in
the gauge $\nabla_i h^i_j=0$.  For the sake of completeness we
also verify directly that the latter path integral coincides
with gauge-fixed path integral of Taylor and Veneziano \cite{A15}.
One can consider this as a by-product or as a consistency
check. Our proof is valid for any dimensionality of space-time.
The proof is given in the next section, where we also discuss
some specific features of quantum gravity. The crucial point is
cancellation of two functional determinant of non-degenerate
second order elliptic operators. This determinants are
identical at a formal level. However, to achieve cancellation
in actual computations one should use gauge-invariant
regularization. This situation is common for all gauge
theories with finite cut-off. A more subtle difficulty is related
to the conformal factor problem \cite{A16}. It appeared that
a kind of complex rotation of the conformal factor is
necessary for gauge independence, as well as for existence of
one-loop path integral.

    In section 3 we compute the one-loop effective action
for quantum gravity on $CP^2$ in a general three-parameter
gauge. Some limiting case of this gauge corresponds to the
Itchinose gauge condition \cite{Ich}. We find no gauge
dependence. We also suggest an explanation of gauge-dependence
of previous result \cite{Ich}. All necessary information
\cite{CP2} about geometry and harmonic expansion on $CP^2$
is collected in the Appendix.

\section{Proof of gauge--independence}

Consider metric fluctuations $h_{ij}$ on a background with
metric tensor $g_{ik}$. Linearised diffeomorphism transformations
act on $h_{ij}$ as follows
\beq h_{ij} \to h_{ij}+(L\xi )_{ij} , \quad
(L\xi )_{ij}=\nabla_i\xi_j +\nabla_j \xi_i \label{eq:(1)}\eeq
where $\nabla_j$ is the covariant derivative with respect to
background metric $g_{ik}$.

Consider gauge condition
\beq (Gh)_j=0 \label{eq:(2)}\eeq
with some operator $G$ depending on background metric $g_{ik}$.
Let the condition (\ref{eq:(2)}) be lineary admissible.
It means that the
operator $GL:\xi \to \xi$ is invertible for all vector fields except
for Killing vectors of $g_{ij}$, and $h_{ij}$ admits unique
decomposition
\beq h_{ij}=\tilde h_{ij}+(Lv)_{ij}, \quad
G\tilde h=0 \label{eq:(3)}\eeq

The Euclidean signature Einstein-Hilbert action reads
\beq S=\frac 1{16\pi G_N} \int d^4x \sqrt g (R+2\Lambda) .
\label{eq:(4)}\eeq
Let the background metric $g_{ik}$ satisfies classical equations of
motion
\beq R_{ik}=-\Lambda g_{ik} \label{eq:(5)}\eeq

Let us introduce a covariant ultralocal inner product in
the space of metric fluctuations
\beq <h,h'>=\int d^4x \sqrt g {h'}_{ik}h^{ik} ,\label{eq:(8)}\eeq
which corresponds to $C=0$ in the De Witt's configuration space
metric.
The path integral measure can be defined by
\beq
\cD h =\frac 1{{\rm vol\ Diff}} \cD_G h . \label{eq:ad}
\eeq
where $\cD_G$ is the Gaussian measure with respect to the inner
product (\ref{eq:(8)}). ${\rm vol\ Diff}=\int \cD \xi$
denotes the volume of the gauge
group. This is an infinite constant which can be put equal to
unity by imposing an appropriate normalization condition.

In a framework of the so-called geometric approach
\cite{A12,A2}  to quantum gravity the path integral
\beq Z(g)=\int {\cal D} h \exp (-S(g,h)) \label{eq:(6)}\eeq
can be evaluated (at one loop) by substituting the decomposition (3)
in (\ref{eq:(6)}), truncating $S(g,h)$ to quadratic order in $h$, and
integrating over $v$. Due to gauge invariance, $S_2(g,h)$ does
not depend on $v$. Hence integration over $v$ gives the
volume of the gauge group which cancelles the corresponding
multiplier in (\ref{eq:ad}). Finally we obtain
\beq Z^{(1)}(g)= \int {\cal D} \tilde h J \exp
(-S_2(g,\tilde h)) , \label{eq:(7)}\eeq
where $J$ is the Jacobian factor due to the change
of variables $h\to \{ \tilde h,v\}$ (see (\ref{eq:(3)}),
 which could be expressed
in a standard
manner \cite{A2}.

We can rewrite
\beq S_2(g,h)=<h,{\bf K}h> \label{eq:(9)}\eeq
with some second order self-adjoint operator ${\bf K}$. Hence at one
loop
\beq Z^{(1)}(g)=(\det {\bf \tilde K})^{-\frac 12} J \label{eq:(10)}\eeq
where ${\bf \tilde K}$ is the operator ${\bf K}$ restricted to
the space defined by  gauge condition (\ref{eq:(2)}).
\newline
{\bf Proposition.} The path integral (\ref{eq:(10)}) does not depend on gauge
condition (\ref{eq:(2)}) provided $G$ is lineary admissible.

To prove the Proposition, consider the gauge $\tilde h=h^\perp$,
where
\beq \nabla_i h^{\perp ik}=0 \label{eq:(11)}\eeq
Due to gauge invariance
\beq \nabla_i ({\bf K}h)^{ik}=0 \label{eq:(12)}\eeq
for arbitrary $h^{ik}$. Consequently ${\bf K}$ can be used as a
projection operator onto the space defined by Eq. (\ref{eq:(11)})
\beq h_{ik}=h_{ik}^\perp +(L\xi )_{ik},
\quad h_{ik}^\perp =({\bf K}^{\perp -1}{\bf K} h)_{ik} ,
\label{eq:(13)} \eeq
where ${\bf K}^\perp$ is the operator ${\bf K}$ restricted to the
space (\ref{eq:(11)}). The Jacobian factor for
this gauge condition (\ref{eq:(11)})
is just \beq J=\det (L^{\dag }L)^{\frac 12} ,
\label{eq:(14)}\eeq
where $L^{\dag }$ is hermitian conjugate to $L$. Note, that
$\det (L^{\dag }L)^{\frac 12}$ is the volume of the gauge
orbit \cite{A2}.

In general, a change of variables from ${\Phi^A}$ with metric
$H_{AB}$ to $\Psi^\alpha$ leads to the Jacobian
\beq J=\det \{ H_{AB} \frac {\delta \Phi^A}{\delta \Psi^{\alpha}}
\frac {\delta \Phi^B}{\delta \Psi^{\beta}} \}^{\frac 12} =
\left( \int {\cal D} \Psi \exp (-<\Phi (\Psi ),\Phi (\Psi )>)
\right) ^{-1} \label{eq:(15)}\eeq
where in the last line we suppose that the change of variables
is linear, $<,>$ is the inner product with respect to the
metric $H_{AB}$. The measure in the last path integral is the
Gaussian one.

Translating the condensed notations in (\ref{eq:(15)})
 to our case we
obtain
\beq J^{-1}=\int {\cal D} \tilde h {\cal D} \xi \exp
(-<\tilde h +L \xi, \tilde h +L \xi >) . \label{eq:(16)}\eeq
We can integrate over $\xi$ in (\ref{eq:(16)}) taking into account
mixing between $\tilde h$ and $\xi$ and obvious relation
$<h^\perp ,L\xi >$$=0$.
\begin{eqnarray}
 J^{-1} & = & (\det L^{\dag } L)^{-\frac 12 } \int {\cal D}
\tilde h \exp (-<\tilde h^{\perp},\tilde h^\perp >)= \nonumber \\
& = & (\det L^{\dag } L)^{-\frac 12 } \int {\cal D}
\tilde h \exp (-<\tilde h, {\bf K}^{\perp -1}
{\bf K}\tilde h >) \label{eq:(17)}\end{eqnarray}
It is easy to see that the last path integral is equal to
\beq (\det {\bf K}^\perp )^{\frac 12}
(\det \tilde {\bf K} )^{-\frac 12} \label{eq:(18)}\eeq
Substituting (\ref{eq:(17)}) and
(\ref{eq:(18)}) in (\ref{eq:(10)}) we obtain
\beq Z^{(1)}(g)=(\det {\bf K}^\perp )^{-\frac 12}
(\det L^{\dag }L)^{\frac 12} \label{eq:(19)}\eeq
One-loop integral (\ref{eq:(10)}) in arbitrary gauge is
equal to the one
in the gauge (\ref{eq:(11)}) and , hence, does not depend on gauge
conditions. This completes the proof.

For the sake of completeness let us verify that on four-sphere
$S^4$ the path integral (\ref{eq:(19)}) coincides with the gauge-fixed
path integral \cite{A15} and with the Hamiltonian path integral
\cite{A8}.

Let us make use of the decomposition
\beq h_{ik}=h_{ik}^{TT}+(L\xi^T)_{ik}+\nabla_i \nabla_k \sigma
-\frac 14 \Delta \sigma +\frac 14 g_{ik}h, \label{eq:(20)}\eeq
where $h^{TT}$ is transverse traceless tensor, $\xi^T$ is
transverse vector, $\sigma$ and $h$ are scalar fields.
Scalars $\sigma$ and $h$ can be decomposed in a sum of
spherical harmonics $Y_l$. The eigenvalues of the Laplace
operator $\Delta$ are $-l(l+3)/r^2$; $r$ is the radius of
$S^4$. For $\sigma$ the sum starts from $l=2$, while for
$h$ - from $l=0$.

The gauge condition (\ref{eq:(11)}) means that $\xi^T=0$,
$h(l=1)=0$. Other scalar harmonics ($l\ge 2$)
are related by the equation
\beq h=-3(\Delta +\frac 4{r^2} )\sigma \label{eq:(21)}\eeq
Let us use the field $\sigma$ as a coordinate in
the configuration space. The Jacobian $\det (L^{\dag} L)$
should be multiplied by another factor arising from
the change of variables $h^\perp \to$
$\{ h^{TT},\sigma (l\ge 2),h(l=0)\}$. After some
algebra one obtaines the total Jacobian factor
\beq J_{tot}=\det_{TV} (-\Delta \delta^k_i +R^k_i)^{\frac 12}
\det_{S(l\ge 2)} \left( (-\Delta +\frac R4 )
(-\Delta +\frac R3)^{\frac 12} \right) \det_{S(l=1)}
(-\Delta )^{\frac 12} , \label{eq:(22)}\eeq
where the first factor refers to transversal vector fields
excluding Killing vectors, the second and the third ones
are computed on scalar harmonics. In our notations the
Ricci tensor $R^k_i$ is negative.

Substituting the fields $\{ h^{TT},\sigma (l\ge 2),h(l=0)\}$
in the Einstein - Hilbert action and truncating to
quadratic order we obtain with the help of equation (\ref{eq:(21)})
\begin{eqnarray}
 S= &\frac 1{32\pi G_N} \int d^4x \sqrt g
[ \frac 12 h^{TT}_{ik}(-\Delta g^{ij}g^{kl}
+R^{ijkl})h^{TT}_{jl}-\nonumber \\
 & -3\sigma (-\Delta +\frac R3 )(-\Delta +\frac R4 )^2
\sigma -\frac 14 h\frac R3 h] \label{eq:(23)}
\end{eqnarray}

As usual, to obtain a convergent path integral the field
$\sigma$ should be rotated to imaginary values
\cite{A13,A16}. After functional
integration we obtain in one-loop order
\beq Z=\det_{TT}
(-\Delta g^{ij}g^{kl}+R^{ijkl})^{-\frac 12}
\det_{VT} (-\Delta \delta^k_i +R^k_i)^{\frac 12}
\det_{S(l=1)} (-\Delta )^{\frac 12} \label{eq:(24)}\eeq
where we omitted contributions of classical action
and global dilatations, $TT$ refers to transverse
traceless tensor fields. This is exactly the result
of Taylor and Veneziano \cite{A15} and of Hamiltonian
quantum gravity on de Sitter space \cite{A8}.

Let us formulate main technical
features of our approach.
 We do not use adapted coordinates (except for
single gauge (\ref{eq:(11)})). We integrate over fields
satisfying arbitrary gauge conditions directly. This
leads to another form of path integral compared
with standard geometric approach.
We construct the projection operator on pure
gauge degrees of freedom using the quadratic form
of the action. This makes the proof more straightforward.
This method can be extended to other gauge theories.
Our result indicates that in a framework of geometric
approach one can consider selfconsistenly the path
integral (\ref{eq:(6)}) without pre-fixing of gauge condition.

    It is more important for quantum gravity that our proof is
formulated in terms of one-loop quantities only. We deal with
determinants of elliptic operators which are unambiguous in any
given regularization. The only formal step is cancellation
of two determinants of ${\bf \tilde K}$. One of them comes
from Gaussian integration in a given gauge, the other is
produced by the Jacobian factor. In principle, these two
determinants can be regularized in different ways. This could
give rise to gauge dependence of one-loop counterterms.
However, this situation is common for all gauge theories.
One-loop counterterms should be gauge-independent if a BRST
invariant regualrization is applied. One can make use of, for
example, the dimensional regularization.

    Another subtle point is related with the conformal factor
problem. To obtain a convergent path integral one should rotate
the trace part of the metric fluctuations to imaginary values.
{}From our proof it is clear that to achieve gauge-independence
one should perform this rotation everywhere simultaneously.
The dimensional regularization is insensitive to overall
constant factor before propagator, thus one can easily overlook
the minus sign produced by the conformal factor rotation.
However, if one studies an expansion of the path integral with
respect to a small parameter which enters gauge conditions,
one can obtain a wrong sign of the contribution of trace part
of graviton, which will no longer cancel gauge-dependent
part of the ghost contribution. In other words, one should
make the conformal factor rotation {\it before \/} expansion
in a small parameter. In the next section we show that the
gauge-dependence of the computations \cite{Ich} can originate
from reversed order of expansion and rotation.

\section{Quantum gravity on $CP^2$ in a general gauge}

    Consider the complex projective space $CP^2$.
All necessary information about geometry of $CP^2$ and the
harmonic expansion is collected in the Appendix. On $CP^2$
there are two invariant covariantly constant rank two tensors,
the metric $g_{ik}$ and the complex structure $J_{ik}$. The
most general linear invariant gauge condition with even
powers of $J$ only reads
\beq G_i(h)=\nabla^j(h_{ji}+\beta J_{jk}J_{il}h^{kl}+
\gamma g_{ji}h_k^k)=0. \label{eq:(3.1)}\eeq
We impose this condition by adding a gauge-breaking term to
the action,
\beq {\cal L}_{gb}=\frac 1{16\pi G_N} \frac 1{2\alpha} G_iG^i.
\label{eq:(3.2)}\eeq
Using an explicit form (\ref{eq:(A.1)}) of the Riemann tensor one can
demonstrate that on this background
the gauge condition studied by Ichinose \cite{Ich}
\begin{eqnarray}
G_i=\nabla^jh_{ji}-\frac 1{2b}\nabla_i h_j^j+\delta
{R_i}^{kjl}\nabla_jh_{kl} \nonumber \\
b-1<<1,\ \ \delta <<G ,\ \ \alpha -1<<1
\label{eq:(3.3)} \end{eqnarray}
 is a particular case of the gauge (\ref{eq:(3.1)}).
In fact, the papers \cite{Ich} deal with small deviations from the
harmonic gauge.

    Here the constants $\alpha$, $\beta$ and $\gamma$ are
arbitrary. For some values of these parameters negative or
zero modes can appear. As was demonstrated by Allen \cite{All},
the infra-red divergencies in graviton and ghost propagators
cancel each other on the de Sitter space. Here we will show
that the ghost contributions to the path integral are canceled
by a part of graviton contribution. This means that the result
\cite{All} can be extended to $CP^2$.

    Total second order on shell action for graviton $h$ and
ghosts $\xi$ and $\bar \xi$ has the following form
\begin{eqnarray}
S=\frac 1{16\pi G_N} \int d^4x \sqrt g [\frac 14 \tilde h^{ik}
(-g_{ij}g_{kl}\Delta +2R_{ijkl})h^{jl}- \nonumber \\
-\frac 12 \nabla^i \tilde h_{ik} \nabla^l\tilde h_l^i+
\frac 1{2\alpha} G_j(h)G^j(h)+ \nonumber \\
+\frac 1{\sqrt \alpha} \bar \xi^i G_i (L\xi )],\quad
\tilde h_{ij}=h_{ij}-\frac 12 g_{ij}h_k^k. \label{eq:(3.4)}
\end{eqnarray}
To produce a damping exponential factor in the path integral
over pure gauge fluctuations the parameter $\alpha$ should be
positive.
The overall factor $(16\pi G_N)^{-1}$ can be absorbed in field
redefinition. For the sake of convenience we use slightly
unusual normalization of the ghost action. It is inessential
in dimensional regularization because it leads to contribution
to the effective action proportional to $\delta^4(0)$.
However, with this normalization cancellation of the ghost
contributions is seen in a more straightforward way.

    Consider first the graviton part. Due to the orthogonality
property (\ref{eq:(A.3)}) and $SU(3)$ invariance of the gauge fixing term
the modes belonging to different $SU(3)$ representations
decouple. One can study contributions of real irreducible
representations of $SU(3)$ separately. We begin with $\mms$
harmonics. Denote by $\phi [\mms ]_q$ corresponding coefficients
in the harmonic expansion (\ref{eq:(A.2)}).
A part of the action (\ref{eq:(3.4)})
depending on these coefficients is
\begin{eqnarray} S[\mms ]=& \nonumber \\
\sum_{m=0}^\infty \sum_{q=1}^{d(m,m+6)}
\phi [\mms ]_q^2 (m^2+8m+15), \label{eq:(3.5)}\end{eqnarray}
where we used expression (\ref{eq:(A.1)}) for the Riemann tensor,
eigenvalues of the Laplace operator (\ref{eq:(A.13)}) and the
transversality equation (\ref{eq:(A.10)}). An overall constant factor
is omitted.

    Due to the orthonormality property (\ref{eq:(A.3)}) the path integral
measure is just
\beq d\mu [\mms ]=\prod_{m,q} d\phi [\mms ]_q. \label{eq:(3.6)}\eeq
Contribution of the $\mms$ modes to the path integral reads
\beq Z[\mms ]=\prod_{m=0}^\infty \left (
\frac {m^2+8m+15}{r^2} \right )^{-d(m,m+6)}. \label{eq:(3.7)}\eeq
Here we restored dependence on the scale factor $r$ defining
size of $CP^2$ after rescaling $g_{ik}\to r^2g_{ik}$.

In the $\mmt$ sector the quadratic form of the action is
represented by a non-minimal differential operator. Hence it
is more convenient to use the functions (\ref{eq:(A.11)}) instead of
standard orthonormal basis.
\begin{eqnarray}
    h_{ij}[\mmt ]  = \cqa Lv_q[\mmt ]_{ij}+ \nonumber \\
 + \cqb {J_i}^k{J_j}^lLv_q[\mmt ]_{kl}, \label{eq:(3.7A)}
\end{eqnarray}
where $v_q[\mmt ]$ are orthonormal vector harmonics.
By substituting (\ref{eq:(3.7A)}) in (\ref{eq:(3.4)})
with the help of (\ref{eq:(A.6)}), (\ref{eq:(A.13)})
and (\ref{eq:(A.14)}) we obtain
\begin{eqnarray}
S(\mmt )=\sum_{q=1}^{20} (c_{1,q}^0 )^2\frac 1{2\alpha}
(3+3\beta)^2
+\sum_{m=1}^\infty \sum_{q=1}^{2d(m,m+3)}
\pmatrix{\cqa \cr \cqb}^T \nonumber \\
\times \pmatrix{
\frac 1{2\alpha}(\lambda +3\beta )^2 &
\frac 1{2\alpha} (\beta \lambda +3)(\lambda +3\beta ) \cr
\frac 1{2\alpha} (\beta \lambda +3)(\lambda +3\beta ) &
\frac 12 (\lambda +3)(\lambda -3) +\frac 1{2\alpha}
(\lambda \beta +3)^2}
\pmatrix{\cqa \cr \cqb}\nonumber \end{eqnarray}
\beq \lambda =m^2+5m+3 \quad \label{eq:(3.8)}\eeq
The tensor harmonics in the right hand side of
eq. (\ref{eq:(3.7A)}) are
not orthonormal. This means that the path integral measure
contains a non-trivial Jacobian factor:
\begin{eqnarray}
d\mu [\mmt ]=J_1 \prod_q dc_{1,q}^0 \times \prod_{m\ge 1,q}
d\cqa d\cqb ,\nonumber \\
  J_1=\prod_q <Lv_q[(0,3)\oplus (3,0),
Lv_q[(0,3)\oplus (3,0)>^{\frac 12} \times\nonumber \\
  \prod_{m=1}^\infty \prod_{q=1}^{2d(m,m+3)} {\det}^\frac 12
\pmatrix{
<Lv^m_q,Lv^m_q> &<Lv^m_q,JJLv^m_q> \cr
<JJLv^m_q,Lv^m_q> &<JJLv^m_q,JJLv^m_q> }=\nonumber \\
  =\left (\frac 3{r^2}\right )^{\frac 12 20} \prod_{m=1}^\infty
\prod_{q=1}^{2d(m,m+3)} \frac 1{r^4} {\det}^\frac 12
\pmatrix{\lambda & 3 \cr 3 & \lambda }=\nonumber \\
  =\left (\frac 3{r^2}\right )^{10}
\prod_{m=1}^\infty \left ( \frac {(\lambda -3)(\lambda
+3)}{r^4} \right )^{d(m,m+3)} , \label{eq:(3.9)}\end{eqnarray}
where $<\ ,\ >$ is the inner product (\ref{eq:(8)}).
We used abbreviated notations
\beq (JJLv)_{ij}={J_i}^k{J_j}^l (Lv)_{kl} .\label{eq:(3.10)}\eeq
Total contribution of gravitons belonging to the
representations $\mmt$ reads
\beq Z[\mmt ]=\left (\frac 3{r^2}\right )^{10}
\prod_{m=1}^\infty \left ( \frac {\lambda
+3\beta}{\sqrt \alpha r^2} \right )^{-2d(m,m+3)} .
\label{eq:(3.11)}\eeq

    In the $(m,m)$ sector let us make use of the harmonics
(\ref{eq:(A.12)}) taking into account trace part of $h_{ij}$
\begin{eqnarray}
h_{ij}[(m,m)]=\sum_{m=2}^\infty \sum_q a_q^m ({J_i}^k
\nabla_k \nabla_j +{J_j}^k \nabla_k \nabla_i )s(m,m)_q+\nonumber \\
  +\sum_{m=1}^\infty \sum_q b_{1,q}^m \nabla_i \nabla_j
s(m,m)_q +\sum_{m=2}^\infty \sum_q b_{2,q}^m {J_i}^k {J_j}^l
\nabla_k \nabla_l s(m,m)_q+\nonumber \\
  +\sum_{m=0}^\infty \sum_q f_q^m g_{ij} s(m,m)_q ,
\label{eq:(3.12)}\end{eqnarray}
$s(m,m)_q$ are orthonormal scalar harmonics. The path integral
measure now reads
\beq d\mu [(m,m)]=J_2 \prod_{m,q} da_q^m db_{1,q}^m db_{2,q}^m
df_q^m. \eeq
The Jacobian factor $J_2$ can be evaluated in the same way as
$J_1$:
\begin{eqnarray}
J_2=\det \pmatrix{ \kappa_1(\kappa_1-\frac 32 ) & -\kappa_1
\cr -\kappa_1 & 4}^{\frac 12 d(1,1)} \times\nonumber \\
\times \prod_{m=2}^\infty \det \pmatrix{
\kappa_m (\kappa_m-3) & 0 & 0 & 0 \cr
0 & \kappa_m (\kappa_m-\frac 32 ) & \frac 32 \kappa_m &
-\kappa_m \cr
0 & \frac 32 \kappa_m & \kappa_m (\kappa_m-\frac 32 ) &
-\kappa_m \cr
0 & -\kappa_m & -\kappa_m & 4}^{\frac 12 d(m,m)} =\nonumber \\
=\left ( \frac {3(\kappa_1-2)\kappa_1}{r^4}
\right )^{\frac 12 d(1,1)} \prod_{m=2}^\infty \left ( 2
\frac {\kappa_m^4 (\kappa_m-3)^2}{r^8} \right )^{\frac 12
d(m,m)}, \nonumber \end{eqnarray}
\beq \kappa_m=m^2+2m=-\Delta (m,m,0,0).\quad \label{eq:(3.13)}
 \eeq
In the last expression we restored dependence of the Jacobian factor
on the scale parameter $r$.

    Substituting the decomposition  (\ref{eq:(3.12)}) in
(\ref{eq:(3.4)}) and performing integration over $a$, $b_1$,
 $b_2$ and $f$ we obtain the following expression for the
contribution of the $(m,m)$ modes to the path integral:
\[
Z(m,m)=J_2 \prod_{m=0}^\infty z_m , \quad
z_0=\left (\frac 2{r^2} \right )^{-\half} , \]
\[
z_1=\left ( -\frac {3(3+3\beta +6\gamma )^2 \kappa_1
(\kappa_1 -2)}{4\alpha r^8} \right )^{-\half d(1,1)} , \]
\beq
z_{m \ge 2}=\left [ -\frac {\kappa_m^4 (\kappa_m-3)^4
(1-\beta)^2 (\kappa_m +\threehalf \beta +\gamma \kappa_m
-\threehalf )^2}{\alpha^2 r^{20}} \right ]^{-\half d(m,m)} .
\label{eq:zmm} \eeq
Note that the eigenvalues in $z_1$ and $z_m$ are negative.
This reflects the conformal factor problem \cite{A16}. To
overcome this difficulty the scalar harmonics $s(m,m)$,
$m\ge 1$ should be rotated to imaginary values during
transition to the Euclidean domain. As it was explained
in the paper \cite{A13}, the Jacobian factor should remain intact.
After the rotation we obtain
\begin{eqnarray}
Z(m,m)=\left (\frac 2{r^2} \right )^{-\half}
\left ( \frac {(3+3\beta +6\gamma )^2 }{4\alpha r^8}
\right )^{-\half d(1,1)} \nonumber \\
\times \prod_{m=2}^\infty
\left [ -\frac { (\kappa_m-3)^2
(1-\beta)^2 (\kappa_m +\threehalf \beta +\gamma \kappa_m
-\threehalf )^2}{2\alpha^2 r^{8}} \right ]^{-\half d(m,m)}
\label{eq:(3.15)}
\end{eqnarray}

    Now we are ready to study the ghost contribution. The
ghosts $\bar \xi$ and $\xi$ can be expanded over the vector
harmonics:
\begin{eqnarray}
\xi_i &=& \sum_{m=0}^\infty \sum_{q=1}^{2d(m,m+3)}
e^m_{1,q} v_i[\mmt ]_q + \nonumber \\
&+& \sum_{m=2}^\infty \sum_{q=1}^{d(m,m)} (
e_{2,q}^m v_i^L(m,m)_q +
e_{3,q}^m v_i^\perp (m,m)_q ) + \nonumber \\
&+& \sum_{q=1}^{d(1,1)} e_{2,q}^1 v_i^L(1,1)_q .
\label{eq:(3.16)}
\end{eqnarray}
The same expression with $e$ replaced by $\bar e$ is valid
for $\bar \xi$. The harmonics $v^\perp (1,1)$ are excluded
since they correspond to the Killing vectors of $CP^2$ and do
not generate any gauge transformations. One can easily check
that these harmonics give zero modes of the ghost action.

    Here we can assume that the basis of vector harmonics
$v_{i,q}$ is orthonormal. Hence the path integral measure
is trivial:
\[
d\mu_{\rm gh}= \prod de_{1,q}^m de_{2,q}^m de_{3,q}^m
d\bar e_{1,q}^m d\bar e_{2,q}^m d\bar e_{3,q}^m \]
The ghost path integral is easily evaluated giving
\begin{eqnarray}
Z_{\rm gh}&=& \prod_{m=0}^\infty \left (
\frac {\lambda +3\beta}{r^2 \sqrt \alpha}
\right )^{2d(m,m+3)} \prod_{m=2}^\infty \left [
\frac {(1-\beta )(\kappa_m -3)}{\sqrt \alpha
r^2} \right ]^{d(m,m)} \nonumber \\
&\times & \prod_{m=1}^\infty \left [
\frac {2\kappa_m+3-3\beta -2\gamma \kappa_m}{\sqrt \alpha r^2}
\right ]^{d(m,m)}.
\end{eqnarray}

    Collecting together all the contributions (\ref{eq:(3.7)}),
(\ref{eq:(3.11)}), (\ref{eq:(3.15)}) and (\ref{eq:(3.7)}) and
neglecting a constant factor $2^\infty$ which can be discarded
in any analytic regularization and which do not contribute to
UV-divergent terms, we obtain
\beq
Z=\prod_{m=0}^\infty \left [ \frac {m^2+8m+15}{r^2}
\right ]^{-d(m,m+6)} \left ( \frac 3{r^2} \right )^{10}
\left ( \frac 2{r^2} \right )^{-\half} .
\label{eq:(3.19)} \eeq
This result is gauge-independent. It coincides with the
earlier calculations \cite{CP2} in a particular gauge.

    Now it is in order to comment on gauge-dependent result
obtained by Ichinose \cite{Ich}. The path integral was evaluated by
expanding in a small parameter which was essentially the same
as $\beta$ here. However, if one expands the path integral
(\ref{eq:zmm})
{\it before \/} conformal rotation in the propagator, one
gets a different sign compared to  expansion of the rotated
expression (\ref{eq:(3.15)}). This leads to doubling of
gauge-dependent term instead of cancellation when one sums
up the corresponding ghost contribution.

    To clarify this point, consider scalar field action
with negative kinetic energy
\beq
S_\varphi =-\int d^4x \varphi (-\Delta +X+\delta {\cal O})
\varphi , \label{eq:ex1}
\eeq
where $X$ and ${\cal O}$ are some operators, $\delta$ is a
small parameter. To obtain convergent path integral one
should rotate $\varphi$ to imaginary values. This gives the
effective action
\begin{eqnarray}
W_{\rm eff} & = & \frac 12 \ln \det (-\Delta +X+
\delta {\cal O}) \nonumber \\
\ & = & \frac 12 \ln \det (-\Delta +X)+
\frac {\delta}2 {\rm tr} \left ( {\cal O}
\frac 1{-\Delta +X} \right ) +O(\delta^2)
\label{eq:ex2}
\end{eqnarray}
If, instead of this, one first expand over $\delta$ and
than perform rotation by changing sign before the
propagator $-\Delta +X$, the effective action takes the
form
\beq
\tilde W_{\rm eff}=\frac 12 \ln \det (-\Delta +X)-
\frac {\delta}2 {\rm tr} \left ( {\cal O}
\frac 1{-\Delta +X} \right ) +O(\delta^2).
\label{eq:ex3}
\eeq
We see, that the sign before $\delta$ is different in
$W_{\rm eff}$ and $\tilde W_{\rm eff}$. The expression
(\ref{eq:ex2}) correspons to the conformal factor
rotation prescription in quantum gravity. The other
expression (\ref{eq:ex3}) presummably corresponds to
the Ichinose procedure. Note, that in dimensional
regularization any constant factor before propagator
can be neglected, and the conformal factor rotation
can be easily overlooked. This "implicit" rotation by
simply neglecting the minus sign before propagator gives
exactly the expression (\ref{eq:ex3}). To obtain correct
result, the vortex term $\delta {\cal O}$ should be
rotated together with the propagator.

\section{Conclusions}

    In this paper we suggested a new proof of
gauge-independence in on-shell one-loop quantum gravity.
This proof is specially tailored for non-renormalizable
theories. It is formulated in terms of functional determinants
of non-degenerate operators which are well-defined in any
suitable regularization scheme. We stress the role of the
conformal factor rotation. We calculated the one-loop
partition function on $CP^2$ in a general $SU(3)$-invariant
background gauge. All gauge-dependent contributions of graviton
fluctuations are canceled by that of ghosts. We also suggested
an explanation of gauge dependence of previous computations
by Ichinose \cite{Ich}.

    Note, that on manifolds with boundaries the problem
of gauge-invariance and gauge-independence becomes much more
complicated (see, e.g. \cite{bou}). It is connected with the
problem of formulation of diffeomorphism invariant boundary
conditions for gravitational fluctuations. As well as in
our case, this has nothing to do with non-renormalizability
of quantum gravity.

Another intersting example of apparent gauge dependence of an
on-shell quantity is provided by non-Abelian gauge theories at
finite temperature. To remove this gauge dependence one should
introduce an infrared regulator which is kept non-vanishing
untill after one goes on-shell \cite{BKS}.

\section*{Appendix: Harmonic expansion on $CP^2$}

In this appendix we collect all necessary formulae \cite{CP2}
related to the harmonic expansion of gravitational
perturbations and ghosts on $CP^2$. Let $J_{ij}$ be
the complex structure, $J_{ij}=-J_{ji}$, $J_i^kJ_k^j=
-\delta_i^j$. If an appropriate normalization is chosen,
the Riemann tensor can be expressed in terms of metric
and complex structure:
\beq R_{ijkl}=\frac 14 (g_{il}g_{jk}-g_{ik}g_{jl})-
\frac 12 J_{ij}J_{kl} +\frac 14 (J_{il}J_{jk}-J_{ik}J_{jl}) .
\label{eq:(A.1)}\eeq
The Ricci tensor is $R_{ik}=-\frac 32 g_{ik}$. $CP^2$ is
a symmetric space, $CP^2=SU(3)/SU(2)\times U(1)$. The isotropy
representation of $AdSU(2)\times U(1)$ in the tangent space
$T_oCP^2$ is $\rOT$, where the first number is the $U(1)$
charge, and the second one is the dimension of $SU(2)$-irrep.
The isotropy representation is reducible as complex
representation and irreducible as real one. The two components
correspond to eigenvalues $\pm i$ of the complex structure
$J$.

For any
homogeneous space $G/H$ a field $\Phi_A$ belonging to
an irreducible representation $D(H)$ can be expanded as \cite{Sal}
\beq \Phi_A (x)=V^{-\frac 12} \sum_{n, \zeta ,q}
\sqrt {\frac {d_n}{d_D}} D^{(n)}_{A\zeta ,q}
(g_x^{-1}) \phi^{(n)}_{q,\zeta }\ , \label{eq:(A.2)}\eeq
where $V$ is the volume of $G/H$, $d_D={\rm dim}D(H)$.
We summarize over representations $D^{(n)}$ of $G$
which give $D(H)$ after reduction to $H$. $\zeta$ labels
multiple components  $D(H)$ in the branching
$D^{(n)}\downarrow H$, $d_n=$$\dim D^{(n)}$. The matrix
elements of $D^{(n)}$ have the following orthogonality
property
\beq \int d^4x \sqrt g
D^{(n) *}_{A\zeta ,q} (g_x^{-1})
D^{(n')}_{A\xi ,p} (g_x^{-1})
=Vd_n^{-1} d_D \delta_{\zeta \xi} \delta_{p q}
\delta_{nn'} \label{eq:(A.3)}\eeq

In the case of $CP^2$ the following representations of
$G=SU(3)$ contribute to the harmonic expansion (\ref{eq:(A.2)}).

    (i){\it Scalar fields.} The representation of
$SU(2)\times U(1)$ is $D(H)=(0,1)$. The representations
of $SU(3)$ which contribute to (\ref{eq:(A.2)}) are
\beq D(SU(3))=(m,m), \quad m\ge 0 \label{eq:(A.4)}\eeq
The $SU(3)$ representations are labeled by the Dynkin
indices, i.e. by the coordinates of a highest weight in
the basis of fundamental weights.

    (ii){\it Vector fields}. The representation $D(H)$
coincides with the adjoint action of $SU(2)\times U(1)$ on
$T_oCP^2$, $D(H)=\rOT$. The following representations
of $SU(3)$ contribute to the harmonic expansion:
\beq \matrix{(1,2),&\quad D(SU(3))=&(m,m),&\quad m\ge 1 \cr
 & & (m,m+3),& \quad m\ge 0 \cr
(-1,2),&\quad D(SU(3))=&(m,m),&\quad m\ge 1 \cr
 & & (m+3,m),& \quad m\ge 0} \label{eq:(A.5)}\eeq
The vector harmonics transforming as $\mmt$ have the following
transversality property
\beq \nabla_jv^j((m,m+3)\oplus (m+3,m))=J_{ij}\nabla^i
v^j((m,m+3)\oplus (m+3,m))=0 \label{eq:(A.6)}\eeq
The two remaining $(m,m)$ harmonics can be identified with
linear combinations of derivatives of scalar harmonics
$s(m,m)$:
\beq v_i^L(m,m)=\nabla_i s(m,m), \quad
v_i^\perp (m,m)=J_{ij}\nabla^j s(m,m) . \label{eq:(A.7)}\eeq

    (iii) {\it Rank two symmetric tensor.} The representation
$D(H)$ is the symmetric tensor product of two vector
representations, $D(SU(2)\times U(1)=$$(0,1)\oplus
[\rTT ] \oplus (0,3)$. The representations in square
brackets combine in a real irreducible representation.
Projections onto irreducible components have the form
\begin{eqnarray}
T_{ik}^{(0,1)} &=& \frac 14 g_{ik} T_j^j, \nonumber \\
T_{ik}^{(2,3)\oplus (-2,3)} &=& \frac 12 [T_{ik}-
J_{ij}J_{kl} T_{jl}], \nonumber \\
T_{ik}^{(0,3)} &=& \frac 12 [T_{ik}+J_{ij}J_{kl} T_{jl}
-\frac 12 g_{ik}T^j_j] \label{eq:(A.8)}\end{eqnarray}
The harmonic expansion for the trace part $T^{(0,1)}$ was
described above. For the other components we have
\beq \matrix{(0,3),&\quad D(SU(3))=&(m,m),&\quad m\ge 1 \cr
 & & (m+3,m),& \quad m\ge 0 \cr
 & & (m,m+3),& \quad m\ge 0 \cr
(2,3),&\quad D(SU(3))=&(m,m),&\quad m\ge 2 \cr
 & & (m,m+3),& \quad m\ge 1 \cr
 & & (m,m+6),& \quad m\ge 0 \cr
(-2,3),&\quad D(SU(3))=&(m,m),&\quad m\ge 2 \cr
 & & (m+3,m),& \quad m\ge 1 \cr
 & & (m+6,m),& \quad m\ge 0} \label{eq:(A.9)}\eeq
Note that every representation $D(G)$ of (\ref{eq:(A.4)}),
(\ref{eq:(A.5)}) and (\ref{eq:(A.9)}) contains corresponding
representation $D(H)$
with unit multiplicity. This means that we can drop out
the index $\zeta$ in (\ref{eq:(A.2)}) and (\ref{eq:(A.3)}).

    The harmonics $T(\mms )$ satisfy the equations
\beq \nabla_iT^i_k (\mms )=\nabla_iJ^{ij}T_{jk}(\mms )=0.
\label{eq:(A.10)}\eeq
The $\mmt$ harmonics can be represented as linear
combinations of vector harmonics:
\beq Lv(\mmt )_{ij} \quad {\rm and}
{J_i}^k{J_j}^l Lv(\mmt )_{kl} \label{eq:(A.11)}\eeq
The two tensor fields in (\ref{eq:(A.11)}) are
independent for $m\ge 1$
and are lineary dependent for $m=0$. The operator $L$ is
given by the eq. (\ref{eq:(1)}),
$Lv_{ij}=\nabla_iv_j+\nabla_jv_i$.
Three $(m,m)$ harmonics can be constructed from scalar
fields:
\beq L\nabla s(m,m),\ \ m\ge 1, \quad Lv^\perp (m,m) \ \
m\ge 2, \quad {J_i}^k{J_j}^lLv^\perp (m,m)_{kl}\ \ m\ge 2.
\label{eq:(A.12)}\eeq
Note that the fields $v^\perp (1,1)$ correspond to the
Killing vectors and do not give rise to tensor fields.

    The eigenvalues of the Laplace operator $\Delta =
\nabla_i\nabla^i$ can be expressed in terms of quadratic
Casimir operators:
\begin{eqnarray}
-\Delta (m_1,m_2;S,Y)=C_2(SU(3))-C_2(SU(2) \times U(1))
\nonumber \\
=\frac 13 (m_1^2+m_2^2+m_1m_2+3m_1+3m_2)-S(S+1)-
\frac 34 Y^2 \label{eq:(A.13)}\end{eqnarray}
where $S$ is the $SU(2)$-spin, $Y$ denotes the $U(1)$
eigenvalues. For example, for the vector component $(\pm 1,2)$
$Y=\pm 1$, $S=\frac 12$. Degeneracies of the eigenvalues are
given by dimensions of corresponding $SU(3)$ representations
\beq d(m_1,m_2)=\frac 12 (m_1+1)(m_2+1)(m_1+m_2+2)
\label{eq:(A.14)}\eeq

\ack{
This work was supported by the Russian Foundation for
Fundamental Studies, grant 93-02-14378.}

\end{document}